\documentclass[sigconf,authorversion,nonacm]{acmart} %

\AtBeginDocument{%
  }

\copyrightyear{2025}
\acmYear{2025}
\setcopyright{rightsretained}
\acmConference[RecSys '25]{Proceedings of the Nineteenth ACM Conference on Recommender Systems}{September 22--26, 2025}{Prague, Czech Republic}
\acmBooktitle{Proceedings of the Nineteenth ACM Conference on Recommender Systems (RecSys '25), September 22--26, 2025, Prague, Czech Republic}\acmDOI{10.1145/3705328.3759309}
\acmISBN{979-8-4007-1364-4/2025/09}

\usepackage{multirow}
\usepackage{makecell}
\usepackage{bm}
\usepackage{subcaption} 

\usepackage{cleveref}

\let\vec\bm

\begin{document}

\title{
Normative Alignment of Recommender Systems via Internal Label Shift
}

\author{Johannes Kruse}
\email{johannes.kruse@jppol.dk}
\orcid{0009-0007-5830-0611}
\affiliation{
  \institution{JP/Politikens Media Group}
  \city{Copenhagen}
  \country{Denmark}
}
\additionalaffiliation{
  \institution{Technical University of Denmark}
  \city{Kongens Lyngby}
  \country{Denmark}
}

\author{Kasper Lindskow}
\email{kasper.lindskow@jppol.dk}
\orcid{0009-0004-6412-0930}
\affiliation{
  \institution{JP/Politikens Media Group}
  \city{Copenhagen}
  \country{Denmark}
}
\additionalaffiliation{
  \institution{Copenhagen Business School}
  \city{Frederiksberg}
  \country{Denmark}
}

\author{Michael Riis Andersen}
\email{miri@dtu.dk}
\orcid{0000-0002-7411-5842}
\affiliation{
  \institution{Technical University of Denmark}
  \city{Kongens Lyngby}
  \country{Denmark}
}

\author{Ryotaro Shimizu}
\email{r2shimizu@ucsd.edu}
\orcid{0000-0002-4841-1824}
\affiliation{
  \institution{University of California San Diego}
  \city{La Jolla}
  \state{California}
  \country{USA}
}
\additionalaffiliation{
  \institution{ZOZO Research}
  \city{Tokyo}
  \country{Japan}
}

\author{Julian McAuley}
\email{jmcauley@eng.ucsd.edu}
\orcid{0000-0003-0955-7588}
\affiliation{
  \institution{University of California San Diego}
  \city{La Jolla}
  \state{California}
  \country{USA}
}

\author{Pierre-Alexandre Mattei}
\email{pierre-alexandre.mattei@inria.fr}
\orcid{0000-0002-1297-908X}
\affiliation{
  \institution{Inria, Université Côte d'Azur}
  \city{Nice}
  \country{France}
}

\author{Jes Frellsen}
\email{jefr@dtu.dk}
\orcid{0000-0001-9224-1271}
\affiliation{
  \institution{Technical University of Denmark}
  \city{Kongens Lyngby}
  \country{Denmark}
}
\additionalaffiliation{
  \institution{Pioneer Centre for Artificial Intelligence}
  \city{Copenhagen}
  \country{Denmark}
}

\renewcommand{\shortauthors}{Kruse et al.}

\begin{abstract}
    Recommender systems optimized solely for user engagement often fail to meet broader normative objectives such as fairness, diversity, or editorial values. We introduce NAILS (Normative Alignment of recommender systems via Internal Label Shift), a simple and scalable method for aligning recommendation outputs with target distributions over item-level attributes (e.g., categories). NAILS modifies the user-conditional item distribution to induce a specified marginal distribution over attributes, leveraging existing user–item preferences without retraining the model. To achieve this, we recast the problem as a form of label shift applied internally within a hierarchical classification framework. Adopting a stakeholder-centric perspective, NAILS enables alignment with global normative goals. Empirically, we show that NAILS consistently improves attribute-level alignment with minimal impact on user engagement, providing a practical mechanism for value-driven recommendation.
    Our code is available at \url{https://github.com/johanneskruse/nails}.
\end{abstract}



\maketitle

\section{Introduction}
Recommender systems play a central role in content delivery across a wide range of domains, including news, entertainment, and e-commerce~\cite{Jannach2019, kruse23_recsys, Shimizu_2024_CVPR, Shimizu_WWW, deldjoo2023_fashionrec_review}. While these systems are typically optimized for user engagement, such optimization may come at the expense of alignment with broader organizational or societal goals~\cite{normalize_2023_first, helberger2019}. In domains like news, for example, purely behavior-driven personalization can reinforce filter bubbles, reduce content diversity, and underrepresent important but less popular content~\cite{Nechushtai2019_newsgatekeepers, Vrijenhoek2021_mission}. This growing concern has motivated research into the normative design of recommender systems---designing recommender systems that are not only effective but also align with normative objectives, such as fairness, diversity, or editorial values~\cite{ge2010_serendipity_coverage, Kaminskas2016, stray2021optimizingforaligningrecommender, lu2020_beyondclicks, Vrijenhoek2022_radio, kruse24_recsys_challenge}.

In this paper, we propose Normative Alignment of recommender systems via Internal Label Shift (NAILS), a lightweight framework that enables alignment of recommendations toward normative objectives. While much of the research on alignment of recommendation has adopted a user-centric perspective~\cite{steck2018, hyunsik2024_calibration, abdollahpouri2023}, i.e., focusing on calibrating recommendations based on individual user interactions, our work takes a stakeholder-centric view, empowering platforms, such as mission-driven news publishers, to align recommendations with global normative objectives while preserving engagement and personalization.
Our contributions in this paper are:
\begin{itemize}
    \item We introduce NAILS, a simple and efficient method for aligning recommendations with global normative objectives across users.
    \item We evaluate NAILS on the EB-NeRD dataset~\cite{kruse24_ebnerd} and demonstrate that it effectively aligns recommendation outputs with diverse normative target distributions.
     \item We motivate future directions for normative alignment, including live deployment, long-term effects, and editorial and policy-driven recommendation systems.
\end{itemize}

\section{Related Work}

Our method is closely related to calibration in recommender systems, a concept introduced by \citet{steck2018}, who argued that recommendation lists should reflect the distribution of content categories in a user's historical preferences. For example, if a user has consumed 80\% romance and 20\% action content, a calibrated system should recommend items in similar proportions. To achieve this, \citet{steck2018} proposed CaliRec, a greedy post-processing algorithm that adjusts recommendation lists to prevent users' minority interests from being overwhelmed by dominant ones---a common side effect of optimizing purely for ranking accuracy.

Subsequent work has further advanced this idea. \citet{seymen2021calibrated} proposed a non-greedy approach by formulating calibration as a constrained optimization problem solved via mixed integer programming. \citet{chen2022DACSR} introduced an end-to-end framework that decouples accuracy and calibration into separate encoders and modifies output distributions to improve diversity and balance. \citet{abdollahpouri2023} formulated the task as a minimum cost flow problem, yielding an efficient and exact solution for calibrated ranking. \citet{hyunsik2024_calibration} proposed LeapRec, a two-phase method combining calibration-disentangled learning during training with a relevance-prioritized re-ranking step for sequential recommendations. These works frame calibration from a user-centric perspective, aligning recommendations with individual users' historical preferences.

In contrast, we adopt a stakeholder-centric perspective, focusing on alignment with broader normative objectives across the user population. While this perspective has been overlooked, a few recent works have explored similar ideas. For example, \citet{Wang-2023-two-sided-calibration} propose Personalized Calibration Targets, aiming to balance user-level interest alignment with a system-level target exposure distribution. Similarly, \citet{Zhao-2020-target-customer} introduce a target customer re-ranking algorithm to adjust the population distribution and composition in the top-$K$ target customers of an item, while preserving recommendation quality. Our approach contributes to this emerging line of work by offering a novel formulation of stakeholder-level calibration as a label shift problem---a perspective that enables a scalable, principled, and model-agnostic mechanism for aligning recommendation distributions with external normative targets. 

\section{Normative Alignment via Internal Label Shift}

Let $p_{\theta}(i \mid u)$ denote the probability that user $u \in \mathcal{U}$ interacts with item $i \in \mathcal{I}$ in a recommender system parameterized by $\theta$. Here, $\mathcal{I}$ denotes the set of candidate items to be ranked, and $\mathcal{U}$ denotes the set of users or their representations, which may include, for example, contextual information and session-specific features. The parameters $\theta$ are typically estimated from historical user–item interaction data. Additionally, consider a probabilistic mapping from items $i \in \mathcal{I}$ to a set of attributes $\mathcal{C}$, represented by $p(c \mid i)$. We assume the joint model
\begin{equation}\label{eq:model-joint}
    p(c, i, u) = p(c \mid i) p_{\theta}(i \mid u) p(u),
\end{equation}
where \( p(u) \) is the marginal distribution of user representations, which could be estimated, e.g., from the empirical distribution of users in our dataset. An important point is that we never actually have to compute it in practice, as we shall see in \Cref{eq:user-dependent-normalization}. The attribute $c \in \mathcal{C}$ could denote, for example, the category of an item. If each item belongs to exactly one category, the mapping becomes deterministic. In this case, if each category $c$ is represented by the set of items belonging to it, then $\mathcal{C}$ forms a partition of $\mathcal{I}$, and the mapping reduces to the indicator function $p(c \mid i) = \vec{1}_c(i)$. We aim to solve a normative alignment problem by replacing the marginal attribute distribution
\( p(c) = \sum_{i \in \mathcal{I}, u \in \mathcal{U}} p(c, i, u) \)
with an alternative target distribution \( \tilde{p}(c) \). Here, \( p(c) \) represents the frequency of user interactions with attribute \( c \) (e.g., categories), and can be estimated from data or computed via marginalization of the joint model \( p(c, i, u) \). For example, in news recommendations, an editor might wish to enforce a desired distribution of news categories that better reflects the newspaper’s editorial values. This problem can be naturally framed as one of adjusting class proportions of the attribute, for which the label shift literature offers principled correction methods \citep{saerens2002adjusting,garg2020unified}. However, our model, c.f.~\Cref{eq:model-joint}, more closely resembles hierarchical classification \citep{koller1997hierarchically,silla2011survey}, where labels are structured in an attribute–item hierarchy. Standard label shift methods assume a non-hierarchical label space and apply the shift at the leaf level—corresponding to individual items—whereas we seek to impose the shift at an internal level, corresponding to attributes.

To solve the normative alignment problem, we consider a new model explicitly constructed to have a normative marginal distribution $\tilde{p}(c)$, defined as
\begin{equation}\label{eq:model-normative}
    \tilde{p}(c,i,u) = \tilde{p}(i, u \mid c) \tilde{p}(c).
\end{equation}
Following \citet{saerens2002adjusting}, we assume that the conditional distribution of users and items given the attribute remains unchanged, i.e., $\forall c \in \mathcal{C}: p(i, u \mid c) = \tilde{p}(i, u \mid c)$. Under this assumption, the normatively aligned model can be expressed as
\begin{equation}\label{eq:shifted-model}
    \tilde{p}(c,i,u) = p(i, u \mid c) \tilde{p}(c) = \frac{p(c \mid i) p_{\theta}(i \mid u) p(u)}{p(c)} \tilde{p}(c).
\end{equation}
The resulting probability that a user interacts with an item under this normative aligned model is then given by
\begin{equation}\label{eq:predictive}
    \tilde{p}(i \mid u) = \frac{\sum_{c \in \mathcal{C}}\tilde{p}(c,i,u)}{\tilde{p}(u)} = \frac{p(u)}{\tilde{p}(u)}  \left( \sum_{c \in \mathcal{C}} \frac{\tilde{p}(c)}{p(c)} p(c \mid i) \right) p_\theta(i \mid u),
\end{equation}
where the user-dependent normalization constant can be computed as
\begin{equation} \label{eq:user-dependent-normalization}
\frac{\tilde{p}(u)}{p(u)} = \sum_{i\in\mathcal{I}} \left( \sum_{c \in \mathcal{C}} \frac{\tilde{p}(c)}{p(c)} p(c \mid i) \right) p_\theta(i \mid u) .
\end{equation}
This correction resembles the ratio proposed by \citet{saerens2002adjusting}, with the additional summation over categories. In practice, it can be implemented by adding the weight $\log \sum_{c \in \mathcal{C}} \frac{\tilde{p}(c)}{p(c)} p(c \mid i) $ to the log probabilities of the model, and passing the result through a softmax layer. This weight can be viewed as a reweighting term that adjusts the recommendations to align with the normative marginal distribution $\tilde{p}(c)$.

Label shift in our setting involves modifying the attribute distribution $p(c)$, which theoretically implies a corresponding shift in the marginal user distribution from $p(u)$ to $\tilde{p}(u)$. However, in practice, we apply the correction while keeping $p(u)$ fixed. As a result, the corrected model does not strictly match the target marginal $\tilde{p}(c)$, but instead offers an approximate solution that preserves the intended directional shift. Despite this mismatch, we find the method to be effective empirically.

To control the influence of normative alignment, we introduce a tunable hyperparameter $\lambda \in [0,1]$ that mixes between the original and normative aligned recommenders, i.e., 
\begin{equation}
    \label{eq:nails-general}
    \tilde{p}_\lambda(i \mid u) := \lambda \tilde{p}(i \mid u) + (1 - \lambda) p_\theta(i \mid u).
\end{equation}

\section{Experimental Setup} 
\label{sec:experimental-setup}
We evaluate NAILS in the context of news recommendations. In this setup, the items are news articles, and the attribute of interest is the category of each article. We assume each article belongs to exactly one category; therefore, $p(c \mid i)$ is an indicator function.

\paragraph{Dataset}
We use the Ekstra Bladet News Recommendation Dataset (EB-NeRD), a large-scale benchmark dataset for news recommendation~\citep{kruse24_ebnerd}. It contains over \(37\) million impression logs collected from more than \(1\) million unique users interacting with over \(125{,}000\) distinct news articles. The articles are labeled with one of eight editorial categories: \textit{entertainment} (23\%), \textit{news} (22\%), \textit{crime} (18\%), \textit{sports} (15\%), \textit{miscellaneous} (9\%) (which includes categories such as \textit{music} and \textit{private finance}), \textit{lifestyle} (7\%), \textit{erotic} (3\%), and \textit{opinion} (3\%). All results are reported on the hidden test set. We use the \(13{,}336{,}710\) impression logs to evaluate ranking performance, and \(200{,}000\) beyond-accuracy samples to assess distributional calibration. We remove the \texttt{auto} category (auto-generated articles) from the candidate list (mainly found in the beyond-accuracy samples), resulting in 155 candidate articles per impression.

\paragraph{Baselines} 
We use NRMS \cite{wu2019-nrms} as our base recommendation model $p_{\theta}(i \mid u)$. It is a widely adopted neural news recommender based on multi-head self-attention, commonly used in recent news recommendation research~\cite{Wu2020MIND, kruse24_recsys_challenge, qi_caum_2022, li_miner2022, Andreea2024}. 

To establish a baseline, we include CaliRec~\cite{steck2018}, a post-hoc reranking method that greedily constructs a top-$K$ list by balancing personalization with alignment to a target distribution over item attributes (e.g., categories). Similar to NAILS, as shown in \Cref{eq:nails-general}, CaliRec includes a tunable hyperparameter $\lambda \in [0,1]$ that interpolates between the original recommender and the normatively aligned objective. We follow the original implementation and apply additive smoothing to prevent zero-probability issues.

\paragraph{Hyperparameter Tuning}
We use Optuna~\cite{optuna_2019} with the Tree-Struc\-tured Parzen Estimator~\cite{watanabe_tpe_tutorial2023} to tune the NRMS model for AUC performance on the EB-NeRD small validation set. The search space is explored over 25 trials. Each model is trained on the training portion ($232{,}887$ samples), with the final 24 hours as hold-out set for early stopping. The validation set contains $244{,}647$ samples. 

Following \citet{wu2019-nrms}, we fix the negative sampling ratio to $K=4$ and use a batch size of 32. News encoders are initialized with the dataset’s open-sourced Word2Vec embeddings~\citep{Mikolov2013} which outperformed randomly initialized embeddings in preliminary experiments. The final model uses the best hyperparameters found during tuning: a learning rate of $10^{-4}$ with the Adam optimizer~\cite{adam_kingma2017}, $\ell^2$-regularization ($\lambda_{\ell^2} = 10^{-4}$), $20\%$ dropout, attention query dimensionality of $100$, and $24$ attention heads each producing a $24$-dimensional output.

After tuning, we merge the training and validation sets and reserve the final day of the validation split for early stopping. 
All experiments are conducted on Amazon EC2 instances using \texttt{g5.xlarge} machines with NVIDIA A10 GPUs.

\paragraph{Evaluation Metrics}
We report both ranking and calibration-aware metrics. To evaluate ranking quality, we use Area Under the Curve (AUC). To access the calibration quality with normative objectives, we compute Kullback–Leibler (KL) divergence at top-$K$ ranks (noted as $@K$) between the distribution of recommended content and the desired target distribution~\cite{steck2018, seymen2021calibrated, chen2022DACSR, abdollahpouri2023, hyunsik2024_calibration}. Lower KL values indicate better alignment with the normative target. Finally, we look at coverage of the candidate list, i.e., we aggregate all recommendations across users and compute the fraction of unique selected articles relative to the full candidate list.

\paragraph{Model Distribution}
To estimate the marginal distribution over categories for the original model, $p(c)$, we aggregate predicted probabilities across all users’ candidate lists, yielding a global estimate of how the model allocates probability mass across categories.

\paragraph{Normative Distributions}
We define two target distributions, $\tilde{p}(c)$, which serve as normative objectives for calibration:
\begin{itemize}
    \item \emph{Editorial distribution:} Computed by aggregating category frequencies across the union of users' candidate lists. Since candidate lists are curated by editors, the resulting distribution approximates editorial intent.
    
    \item \emph{Uniform distribution:} A distribution that promotes equal representation across all categories, defined over the set of unique categories present in the candidate lists.
\end{itemize}

\section{Results and Discussions}
\label{sec:results}
\subsection{Normative Distributions}
We first evaluate how well the methods align the distribution of recommended articles with a given normative distribution. This analysis is conducted on the top-\(10\) articles selected per user from the beyond-accuracy test set (see \Cref{sec:experimental-setup}). We consider two variants of NAILS: deterministic (NAILS-det) selecting articles based on highest-probability scores, and a stochastic (NAILS-stoch) sampling articles according to their probability scores.

\paragraph{Alignment Effectiveness}
\Cref{tab:calibration-results} reports KL-divergence and coverage for each method across varying values of \( \lambda \). As \( \lambda \) increases, KL-divergence consistently decreases for all methods, indicating improved alignment with the target normative distribution. 
However, for the editorial distribution, KL-divergence begins to increase again at higher \( \lambda \) values, suggesting that \( \lambda \) serves as a tunable parameter that balances alignment strength and ranking quality. CaliRec achieves the lowest KL@10 overall, which is expected given its greedy optimization procedure that explicitly minimizes KL-divergence. In contrast, NAILS nudges the distribution toward the target but does not directly optimize for KL, and therefore does not guarantee or force exact alignment.

As expected, NAILS-stoch achieves full coverage of the candidate pool, as articles are sampled probabilistically based on their aligned scores. While CaliRec's coverage steadily decreases, NAILS-det increased as \(\lambda\) increases. This suggests that, with appropriate tuning, NAILS-det can better balance distributional alignment and content diversity compared to strictly optimization-based approaches.

\begin{figure}
    \centering
    \begin{subfigure}{\linewidth}
        \centering
        \includegraphics[width=\linewidth]{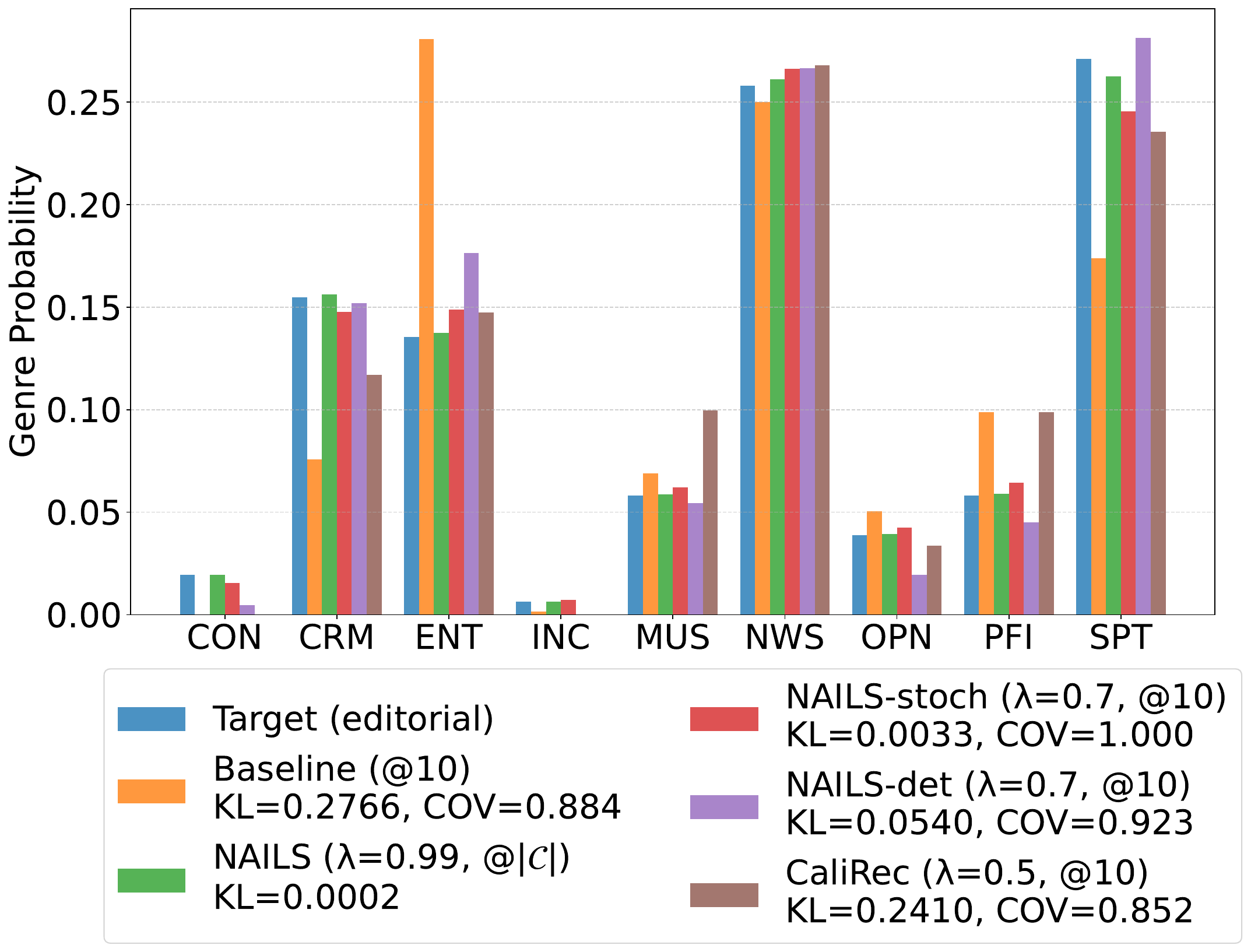}
        \caption{Editorial distribution}
        \label{fig:bar_distribution_best_editorial}
    \end{subfigure}
    \begin{subfigure}{\linewidth}
        \centering
        \includegraphics[width=\linewidth]{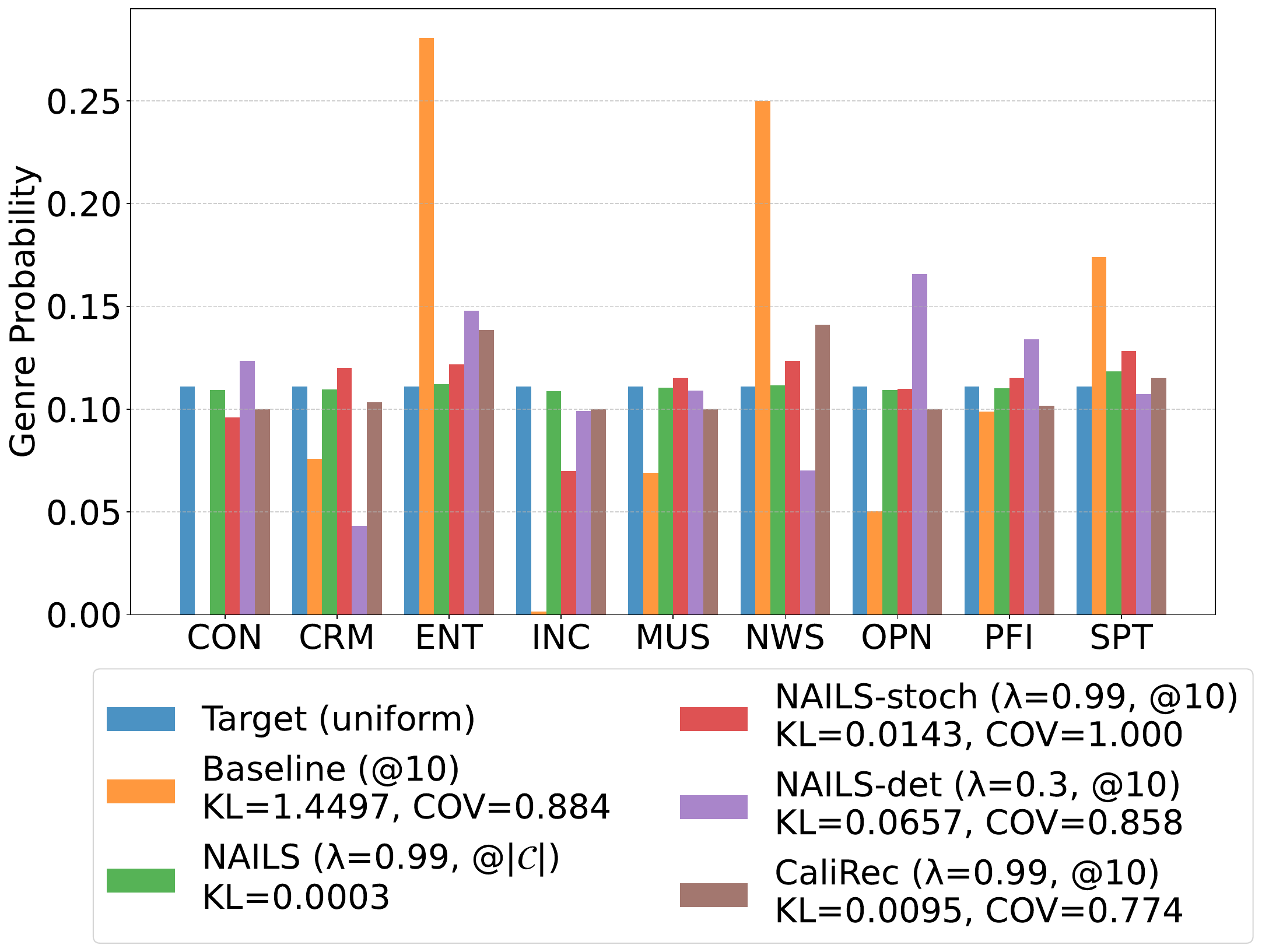}
        \caption{Uniform distribution}
        \label{fig:bar_distribution_best_uniform}
    \end{subfigure}
    \caption{Category distribution among the top-10 recommended articles for alignment toward the Editorial distribution (a) and the Uniform distribution (b). We report KL-divergence (KL, $\downarrow$), aggregated across all users. KL@$|\mathcal{C}|$ refers to the predictive distribution over the full candidate list. 
    The categories are consumption (CON), crime (CRM), entertainment (ENT), income (INC), music (MUS), news (NWS), opinion (OPN), private finance (PFI), and sports (SPT).}
    \label{fig:distribution-comparison}
    \Description[short]{long}
\end{figure}

\paragraph{Distributional Comparison}
\Cref{fig:distribution-comparison} visualizes the aggregated category distributions produced by the best-performing configuration of each method (selected from \Cref{tab:calibration-results}), shown alongside the target distribution, the uncalibrated baseline, and NAILS at top-\( \mathcal{C} \). These results show that NAILS effectively aligns with the stakeholder-defined distribution at the system level when aggregating recommendations across users. The comparison also highlights that the effectiveness of a given method depends on the alignment objective. For example, NAILS performs better under the editorial distribution, while CaliRec achieves closer alignment under the Uniform distribution.

\begin{table*}
\centering
\caption{KL-divergence (KL, $\downarrow$) and coverage (COV) for the top-10 recommended articles under different alignment methods. 
We report KL as the mean KL-divergence computed per user.}
\label{tab:calibration-results}
\scalebox{0.98}{
\begin{tabular}{@{}lcccccccccccc@{}}
\toprule
& \multicolumn{6}{c}{\textbf{Editorial distribution}} & \multicolumn{6}{c}{\textbf{Uniform distribution}} \\
\cmidrule(lr){2-7} \cmidrule(lr){8-13}
& \multicolumn{2}{c}{\textbf{NAILS-stoch}} 
& \multicolumn{2}{c}{\textbf{NAILS-det}} 
& \multicolumn{2}{c}{\textbf{CaliRec}} 
& \multicolumn{2}{c}{\textbf{NAILS-stoch}} 
& \multicolumn{2}{c}{\textbf{NAILS-det}} 
& \multicolumn{2}{c}{\textbf{CaliRec}} \\
\cmidrule(lr){2-3} \cmidrule(lr){4-5} \cmidrule(lr){6-7}
\cmidrule(lr){8-9} \cmidrule(lr){10-11} \cmidrule(lr){12-13}
\textbf{$\lambda$} & KL@10 & COV@10 & KL@10 & COV@10 & KL@10 & COV@10 & KL@10 & COV@10 & KL@10 & COV@10 & KL@10 & COV@10 \\
\midrule
0.0 & 2.060 & 100 & 3.802 & 88.4 & 3.802 & 88.4 & 3.889 & 100 & 4.872 & 88.4 & 4.872 & 88.4 \\
0.1 & 2.052 & 100 & 3.707 & 89.7 & 1.630 & 90.3 & 3.670 & 100 & 3.905 & 88.4 & 2.774 & 88.4 \\
0.2 & 2.046 & 100 & 3.626 & 92.3 & 0.735 & 88.4 & 3.473 & 100 & 3.318 & 87.1 & 1.442 & 86.5 \\
0.3 & 2.036 & 100 & 3.555 & 92.3 & 0.574 & 85.8 & 3.290 & 100 & \textbf{3.160} & 85.8 & 0.481 & 83.2 \\
0.4 & 2.030 & 100 & 3.499 & 92.9 & 0.512 & 83.9 & 3.129 & 100 & 3.253 & 84.5 & 0.084 & 80.6 \\
0.5 & 2.028 & 100 & 3.460 & 92.9 & \textbf{0.496} & 85.2 & 2.983 & 100 & 3.414 & 80.6 & 0.035 & 78.7 \\
0.6 & 2.026 & 100 & 3.434 & 92.9 & 0.502 & 84.5 & 2.858 & 100 & 3.602 & 80.0 & 0.029 & 77.4 \\
0.7 & \textbf{2.025} & 100 & \textbf{3.422} & 92.3 & 0.527 & 83.9 & 2.745 & 100 & 3.783 & 79.4 & 0.029 & 77.4 \\
0.8 & 2.026 & 100 & 3.425 & 92.3 & 0.562 & 83.9 & 2.650 & 100 & 3.916 & 76.8 & 0.028 & 77.4 \\
0.9 & 2.027 & 100 & 3.441 & 92.3 & 0.572 & 85.2 & 2.564 & 100 & 3.994 & 74.2 & 0.028 & 77.4 \\
0.99& 2.031 & 100 & 3.464 & 91.6 & 0.571 & 81.9 & \textbf{2.508} & 100 & 4.034 & 73.5 & \textbf{0.028} & 77.4 \\
\bottomrule
\end{tabular}
}
\end{table*}

\subsection{Ranking Performance}
\begin{figure}
    \centering
    \includegraphics[width=\linewidth]{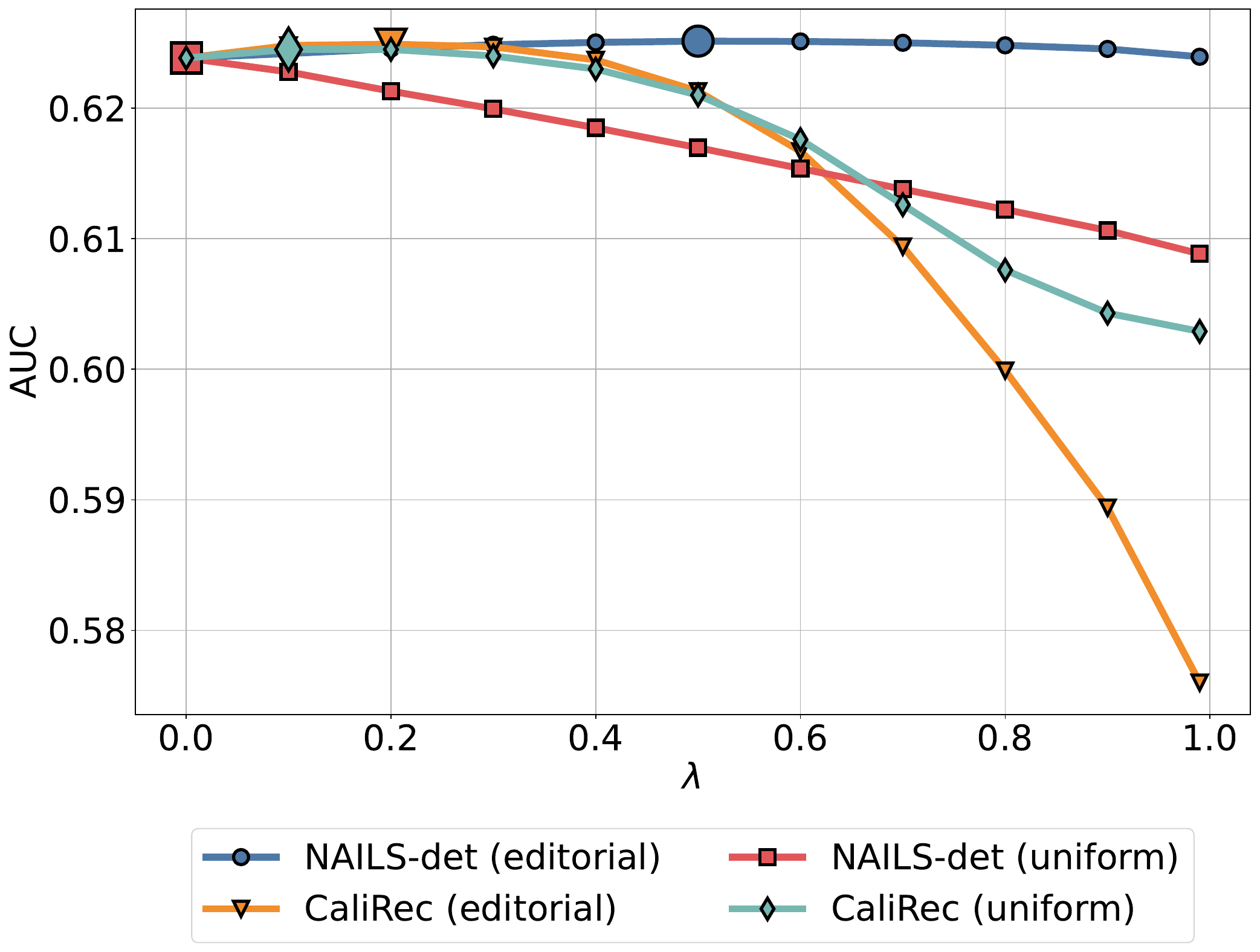}
    \caption{AUC ($\uparrow$) across varying calibration strengths (\( \lambda \)). Larger markers indicate the best performance.}
    \label{fig:ranking_der_lambda}
    \Description[short]{long}
\end{figure}

We now turn to the effect of alignment on ranking quality, as measured by AUC. This analysis is conducted on the hidden test set used for evaluating ranking performance. We focus exclusively on deterministic methods, as our goal is to assess the effect of alignment in a stable and reproducible setting. \Cref{fig:ranking_der_lambda} illustrates how different alignment strengths (\( \lambda \)) influence AUC under different target distributions. We observe that the choice of normative target distribution significantly affects ranking performance. In particular, certain distributions can even lead to improvements. For example, under the editorial distribution, applying alignment improves AUC for both CaliRec and NAILS, with NAILS achieving slightly higher performance overall. In contrast, alignment to the Uniform distribution generally reduces AUC, reflecting the challenge of balancing strict fairness with user relevance. These results highlight an important trade-off between ranking performance and alignment with normative goals. They also underscore the importance of carefully designing target distributions. Depending on the application, alternative targets could be constructed to encourage specific properties, such as greater content diversity, while minimizing the loss in engagement metrics.

\section{Conclusion and Future Work}
This paper proposed Normative Alignment of Recommender Systems via Internal Label Shift (NAILS), a novel method for aligning recommendation distributions with normative objectives. Our experiments demonstrate that NAILS effectively aligns the category distribution of recommendations. Moreover, we find that the choice of target distribution can significantly influence both ranking performance and calibration outcomes. Compared to the established calibration method CaliRec, NAILS offers a lightweight and highly efficient alternative that avoids solving a greedy optimization problem during inference. More broadly, we do not claim universal superiority for any one method. The notion of optimality is inherently context-dependent and depends on the specific deployment goals, whether prioritizing ranking metrics like AUC, KL-divergence, or coverage. Our intention is to highlight the trade-offs embodied by different approaches. While the results are promising, future work is needed to explore the long-term effects of applying normative objectives in live environments, including their impact on user engagement, diversity, and system dynamics over time. Furthermore, although most prior alignment work has centered on a user-centric perspective, we argue that extending calibration to align with global normative goals across users presents a promising and necessary direction for future research, particularly in editorial and policy-driven recommendation systems. Ultimately, incorporating normative alignment frameworks like NAILS could play a key role in building responsible recommendation systems that are aligned with broader normative objectives, such as societal values.

\begin{acks}
We would like to extend our gratitude to and acknowledge our employers and funding bodies, including Ekstra Bladet, JP/Politikens Media Group, Technical University of Denmark, Copenhagen Business School, Innovation Foundation Denmark (grant number 1044-00058B), and the Platform Intelligence in News-Project (grant number 0175-00014B). 
\end{acks}

\balance %
\bibliographystyle{ACM-Reference-Format}
\bibliography{main}

\end{document}